\documentclass[prb,showpacs,superscriptaddress,twocolumn,a4paper]{revtex4}

\usepackage{graphicx}
\usepackage{amsmath}
\usepackage{amssymb}
\usepackage{hyperref}\hypersetup{colorlinks=true,linkcolor=blue,citecolor=blue,pdfstartview=FitH}

\newcommand{\figwidth}{0.95\columnwidth}

\DeclareMathOperator{\Tr}{Tr}

\begin{document}
\title{Quantum quench dynamics of the Coulomb Luttinger model}
\author{N.~Nessi}
\author{A.~Iucci}
\affiliation{Instituto de F\'{\i}sica La Plata (IFLP) - CONICET and Departamento de F\'{\i}sica,\\
Universidad Nacional de La Plata, CC 67, 1900 La Plata, Argentina}

\begin{abstract}
We study the non-equilibrium dynamics of the Luttinger model after suddenly turning on and off the bare Coulomb interaction between the fermions. We analyze several correlation functions such as the one particle density matrix and vertex correlations, its finite time dynamics and the stationary state limit.  Correlations exhibit a non-linear light cone effect: the spreading of the initial signal accelerates as a consequence of the quantum nature of the excitations, whose peculiar dispersion of plasmonic type in 1D gives rise to a logarithmic divergence in the group velocity at $q=0$. In addition we show that both the static and dynamic stationary state correlations can be reproduced with a simple generalised Gibbs ensemble despite the long-range character of the interactions which precludes the application of the Lieb-Robinson bounds. We propose a suitable experimental setup in which these effect can be observed based on ultracold ions loaded on linear traps.

\end{abstract}

\maketitle
\section{Introduction}

Recent experimental progress in the field of ultracold atomic gases loaded in optical lattices~\cite{bloch08_cold_atoms_optical_lattices_review} opened new perspectives in the research of isolated quantum systems out of equilibrium~\cite{greiner02_fast_tunnability,kinoshita06_non_thermalization,chenau12_light_cone_experimental_bosons,trotzky12_relaxation_isolated_bose_gas,gring12_pre-thermalization_isolated_bose_gas}. In particular it has made possible to study the evolution of many-body strongly correlated closed systems after a quantum quench, \textit{i. e.}, an abrupt change of one of the parameters of the system's Hamiltonian. Such out-of-equilibrium setups can be cleanly realized in the experiments with cold atoms due to the fine control over the effective parameters of the system, such as external fields and interactions between the atoms, even in real time. In parallel, there has been an intense theoretical effort in the study of the out of equilibrium dynamics in many-body systems (see Ref.~\onlinecite{polkovnikov11_nonequilibrium_dynamics} for a review). The investigations have focused mostly on the features of the steady state after a quantum quench, the conditions under which it is described by a (generalized) Gibbs ensemble and the existence of metastable states at intermediate times and whether they relax for longer times. Other relevant and interesting aspects that arise are related to the dynamics after the quench in connection to the so called `light-cone effect'~\cite{lieb72_bound_spin_chain,calabrese06_quench_CFT}. 

These aspects of the non-equilibirum dynamics after a quantum quench have been investigated in various specific models such as the quantum Ising chain~\cite{barouch70_quench_ising_chain,sengupta04_quench_QCP,rieger11_semiclassical_theory_qp_quench_ising,igloi12_quench_dynamics_quasicrystals,foini11_quench_FDT}, one dimensional (1D) bosonic models\cite{kollath07_quench_BH,rigol07_generalized_gibbs_hcbosons,iyer12_quench_lieb-liniger,mossel12_quench_lieb_liniger}, the sine-Gordon~\cite{iucci10_quench_sine_gordon,gritsev07_spectroscopy_quench,degrandi11_quench_qcp_sg_model}, and the Luttinger model (LM)~\cite{cazalilla06_quench_LL,iucci09_quench_LL}. General results have been obtained from theoretical investigations involving conformal field theory (CFT)~\cite{calabrese05_evolution_entanglement_1d,calabrese06_quench_CFT,calabrese07_quench_CFT_long}. The LM, and other closely related 1D models, are among the most attractive candidates for studying 1D systems out of equilibrium due to the availability of a manageable exact solution from which one can extract correlation functions in closed form. On the other hand, the (short-range) LM is the fixed point of a large class of 1D gapless systems, called Luttinger liquids; the question of under what conditions this universality property is extensive to the non-equilibrium domain constitutes an open problem. \textit{A priori} we cannot expect the results obtained for the LM out of equilibrium to hold for more general systems, since couplings that are irrelevant in equilibrium, might become relevant when the system is driven out of equilibrium~\cite{mitra11_quench_mode_coupling}, at least for long times. From another point of view, the dynamics after a quenching process will excite high-energy states that are not well described by the LM. Still, in Refs.~\onlinecite{karrasch12_ll_universality_quench,panfil12_lieb_liniger_out_equilibrium_luttinger,pollman12_ll_linear_quench_universality} it has been shown that some characteristics of the long-time dynamics after a short-range interaction quench of certain 1D systems are universal in the Luttinger liquid sense.

So far, theoretical investigations on quench dynamics have focused primarily on systems that interact via contact or, more generally, short-range potentials, \textit{i.e.}, that decay faster than the inverse of the distance between particles. The origin of this bias can be found in the short-range character of the couplings in the prevailing experimental setups that motivated the field in recent years~\cite{greiner02_fast_tunnability,kinoshita06_non_thermalization}. Even in the more traditional condensed matter systems, the Coulomb interaction between electrons is very often screened, particularly in non-isolated 1D systems~\cite{schulz83_screening_coulomb_1d}, giving rise to an effective short-range interaction description in most of the cases. However, this does not exhaust the possibilities of studying out of equilibrium systems, and in particular the consequences of a long-range potential seems to be unexplored, though some results are available in a Hubbard chain with long-range hopping~\cite{kollar08_generalized_gibbs_hubbard1r}. A concrete realization of particles interacting via a long-range potential are linear ion traps~\cite{haffner08_ions_report}. These systems, as their neutral atomic counterparts, turn out to be highly tuneable and are able to maintain coherence for long times, becoming thus ideal candidates for studying out of equilibrium phenomena. 

Regarding the models used in the past to account for long-range potentials, it is specially interesting the case of the LM with Coulomb interactions. Its equilibrium properties have been studied in a series of theoretical works~\cite{schulz93_wigner_crystal_1d,wang01_coulomb_ll,gindikin02_dynamical_correlations_coulomb_ll}, leading to some well known results, such as the prediction of the Wigner crystal phase formation of 1D electrons, which has been recently measured in isolated carbon nanotubes~\cite{deshpande08_wigner_crystal_1d_carbon_nanotubes}. From a perturbative perspective, the fast exponential decay of the (fourier transformed) long-range potential with the momentum exchanged between colliding particles ensures that the sudden connection of interactions is likely not to scatter particles to high-energy states. Therefore, we expect the Coulomb LM to faithfully capture universal properties of more general long-range systems.

In this work we shall consider the non-equilibrium dynamics after a quantum quench of the LM with long-range interactions. We pay most of the attention to the case of a sudden turning on of the Coulomb repulsion between the fermions starting from an arbitrary short-range interacting initial state, though we briefly discuss the opposite case, of a sudden turning off of the Coulomb potential. We analyze in detail the finite time dynamics after the quench, and show, in particular, that correlations spread with increasing velocity, allowing for a propagation of information faster than in systems with short-range interactions. 

The rest of the paper is organized as follows: in section \ref{sec:model}, we  introduce the LM and describe its exact solution in terms of bosonic quasiparticles (QPs). The main results of the work are presented in section~\ref{quench}, where we describe the stationary state correlations and the finite time dynamics. We consider the case where the long-range interaction between the fermions is suddenly switched on and the reverse situation, that is, when the interaction in suddenly switched off. In section~\ref{exp} we proposed an experimental realization of the studied system using ultracod ions loaded in electromagenetic cavities. Finally, we summarize the main results in section~\ref{sec:conclusions}.

\section{The Luttinger model and the bosonization solution}\label{sec:model}

The LM describes a system of interacting fermions in one dimension~\cite{tomonaga50_1D_electron_gas,luttinger63_model,mattis65_lieb_luttinger_model,giamarchi04_book_1d}. The key simplification of the model is the assumption of a linear dispersion relation for the free fermions, which induces a clear separation between right and left moving particles. The LM Hamiltonian is
\begin{equation}\label{ham_lm}
H_{LM}=H_0+H_2+H_4,
\end{equation}
where
\begin{equation}
H_0=\sum_{q,r=R,L} v_F q :c^{\dagger}_{q,r} c_{q,r}:,
\end{equation}
is the free piece of $H_{LM}$. Here $c^{\dagger}_{q,r}$ and $c_{q,r}$ are fermionic creation and annihilation operators at momentum $q$ and $v_F$ is the Fermi velocity. The index $r$ labels the \emph{chirallity} of the particles, denoting left (L) and right (R) moving fermions. The fermionic normal ordering denoted by $:\ldots:$ is needed to remove from the expectation values the infinite contribution arising from the fact that the ground state is a Dirac sea~\cite{haldane81_luttinger_liquid}, namely, a state where all single-particle fermion levels with $q<0$ are occupied.

The strength of the scattering can be parametrized using two functions, $g_2(q)$ and $g_4(q)$, that are related to processes that exchange fermions between the two branches of the spectrum and those that leave the fermions on its original branch respectively. In terms of these, the interacting piece of $H_{LM}$ reads
\begin{eqnarray}
  H_2 &=& \frac{1}{L}\sum_q g_2(q):\rho_{qR}\rho_{qL}:,\\
  H_4 &=& \frac{1}{2L} \sum_q g_4(q):\rho_{qR}\rho_{-qR}+\rho_{qL}\rho_{-qL}:,
\end{eqnarray}
where we have defined densities in momentum space as $\rho_{qr}=\sum_k :c^{\dagger}_{k-q,r} c_{kr}:$. In what follows we will take $g_2(q)=g_4(q)=V(q)$ as the Fourier transform of the two body interaction potential between the fermions $V(x)$.

The elementary excitations of the system are bosonic QPs describing low energy collective density modes of the system~\cite{mattis65_lieb_luttinger_model,giamarchi04_book_1d}. In order to see this, we first note that the density operators $\rho_{qr}$ obey the following commutation rules:
\begin{equation}\label{commut_dens}
[\rho^{\dagger}_{qr},\rho_{r'q'}]=-r\delta_{rr'}\delta_{qq'}n_q,
\end{equation}
where we have defined the integer $n_q=\frac{L q}{2\pi}$. The special algebra of Eq. (\ref{commut_dens}) can be transformed into the usual bosonic commutation relations by introducing the following operators:
\begin{equation}\label{bosons}
b(q)= \frac{1}{\sqrt{|n_q|}}\left\{
                              \begin{array}{ll}
                                \rho_{qR}, & \hbox{for $q>0$,} \\
                                \rho_{qL}, & \hbox{for $q<0$.}
                              \end{array}
                            \right.
\end{equation}
Notice that the $q=0$ components (the zero modes) require a separate treatment since $\rho_{0r}=N_r$ is the deviation, relative to the ground state, in the number of fermions of a given chirality. It is customary to introduce the combinations
\begin{equation}
N=N_R+N_L,\qquad  J=N_R-N_L.
\end{equation}
In therm of these, and of the bosonic operators introduced in Eq. (\ref{bosons}) the Hamiltonian of the LM writes
\begin{eqnarray}
  H_0 &=& \sum_{q\neq0}v_F|q|b^{\dagger}_q b_q+\frac{\pi v_F}{2L}(N^2+J^2), \\
  H_2 &=& \frac{1}{2}\sum_{q\neq0}V(q)|q|[b_q b_{-q}+b^{\dagger}_q b^{\dagger}_{-q}] \\
       &&+\frac{\pi V(0)}{2L}(N^2-J^2) \\
  H_4 &=& \sum_{q\neq0}V(q)|q|b^{\dagger}_q b_q\\
       &&+\frac{\pi V(0)}{2L}(N^2+J^2).
\end{eqnarray}

Ignoring the zero mode sector (that will not contribute in the $L\rightarrow \infty$ limit), we can diagonalize the above bosonic Hamiltonian by means of a Bogoliubov transformation by introducing new operators $a_q$ and $a^{\dagger}_q$ as follows:
\begin{equation}
\left(
  \begin{array}{c}
    a_q \\
    a^{\dagger}_{-q}\\
  \end{array}
\right)=\left(
          \begin{array}{cc}
            \cosh\theta(q) & \sinh\theta(q) \\
            \sinh\theta(q) & \cosh\theta(q) \\
          \end{array}
        \right)\left(
  \begin{array}{c}
    b_q \\
     b^{\dagger}_{-q}\\
  \end{array}
\right),
\end{equation}
where the parameter of the transformation $\theta(q)$ satisfies the relation
\begin{equation}\label{theta}
\tanh2\theta(q)=\frac{V(q)}{2\pi v_F+V(q)}.
\end{equation}
In terms of the new operators the Hamiltonian is rendered diagonal:
\begin{equation}
H_{LM}=H_0+H_2+H_4=\sum_{q\neq0}\epsilon(q)a^{\dagger}_q a_q+\mathrm{zero\ mode\ terms},
\end{equation}
where $\epsilon(q)=v_F|q|\sqrt{1+2\frac{V(q)}{\pi v_F}}$ is the dispersion relation of the bosonic QPs. This defines the equilibrium solution of the LM, with which all correlation functions can be calculated. This is of particular importance since the equilibrium LM is the renormalization group fixed point of a large class of gapless one dimensional models~\cite{haldane81_luttinger_liquid,haldane81_effective_harmonic_fluid_approach}, called Luttinger liquids.

For systems with short-range interactions the LM calculations give rise to the characteristic power law decays of the correlations~\cite{haldane81_luttinger_liquid,giamarchi04_book_1d}. It is worth clarifying that in this work we shall call short-range potential to any potential $V(x)$ whose Fourier transform $V(q)$ is finite at $q=0$, falling in this category potentials that are usually termed as long-ranged, such as the dipole-dipole interaction $V(x)\sim x^{-3}$. In such cases, the asymptotic form of several correlation functions is controlled by the value of the potential at $q=0$, through the so-called Luttinger parameter $K=(1+\frac{V(0)}{\pi v_F})^{-1/2}$ and the renormalized velocity $v=v_F\sqrt{1+\frac{V(0)}{\pi v_F}}$. For systems in which the Coulomb potential is unscreened ($V(q)$ diverging logarithmically for $q\rightarrow 0$, see Section~\ref{sec:on}) the LM predicts the existence of the analogous of a Wigner crystal (the crystalline phase of electrons) in 1D. In particular, as was shown by Schulz~\cite{schulz93_wigner_crystal_1d}, the $4k_F$ part of the density-density correlation function decays as $e^{-C \log^{1/2}(x)}$, much slower than a power law, signaling the stability of a phase in which the electrons fluctuate around equidistant lattice sites.

\section{Interaction quenches in the Luttinger model}\label{quench}

Even though general quenches between quadratic Hamiltonians such as the LM can be fully solved~ \cite{cazalilla06_quench_LL,iucci09_quench_LL}, we shall focus on a specific situation in which the range of the interaction is suddenly modified. Let us assume that at $t=0$ the system is prepared in the ground state $|GS\rangle_i$ of the Hamiltonian $H_i$, defined through Eq. (\ref{ham_lm}) with an interaction potential $V_i(q)$. The Hamiltonian that henceforth dictates the temporal evolution, $H_f$, has a different interaction potential $V_f(q)$. At the level of the bosonic representation, both Hamiltonians can be put in diagonal form by means of canonical Bogoliubov transformations characterized by different parameters $\theta_i(q)$ and $\theta_f(q)$, related to the respective potentials by Eq. (\ref{theta}). The solution of the problem is then reduced to a change of basis consisting of a series of nested Bogoliubov transformations performed in order to find the temporal dependence of the operators that diagonalize the bosonic version of Hamiltonian $H_f$. We show the details of this procedure in the Appendix \ref{app_formal_sol}. We shall mostly consider a simple protocol, according to which the interactions described by $H_2$ and $H_4$ are suddenly changed from short to long-range, being the case of an initial non-interacting Hamiltonian ($g_2(q)=g_4(q)=0$) a specific example. We shall deliver most of the attention to this protocol, and briefly analyze the reverse one in Sec.~\ref{toff}.

\subsection{Turning on Coulomb interactions}\label{sec:on}

Consider that initially we have a one-dimensional gas of electrons interacting via a short-range potential, whose subsequent temporal evolution occurs with the electrons interacting via the Coulomb potential. Since the inclusion of a finite but short range initial potential would only renormalize the value of the Luttinger parameter $K_i$, we shall only consider a contact potential $V_i(x)=s\delta(x)$, which induces an initial Luttinger parameter $K_i=(1+\frac{s}{\pi v_F})^{-1/2}$. We can always recover the limit of a non-interacting initial state by setting $K_i=1$. For the Coulomb potential we take the form $V_f(x)=\frac{e^2}{\sqrt{x^2+d^2}}$ with Fourier transform $V_f(q)=2e^2 K_0(qd)$, where $K_0(\zeta)$ is the modified Bessel function of the second kind of order zero and $d$ is a short-distance cutoff. This phenomenological form has been proposed to effectively describe the unscreened Coulomb potential on semiconductor wires~\cite{gold90_semiconductor_wire}, in which case $d$ is associated to the transverse dimensions of the wire. The spectrum of the bosonic QPs of the Hamiltonian is of palsmonic type in 1D:
\begin{equation}\label{spectrum}
\epsilon_f(q)=|q|v_F\sqrt{1+2 g K_0(qd)}\sim |q|\log^{1/2}\left(\frac{1}{qd}\right)
\end{equation}
with $g=\frac{e^2}{\pi v_F}$, and the logarithmic form is the approximate result for small $q$. The energy of the QPs goes to $0$ for $q\rightarrow 0$, nevertheless its group velocity $v_f(q)=\frac{d\epsilon_f(q)}{dq}$ diverges as $v_F\sqrt{1- 2 g \log(qd)}$ for $q\ll d^{-1}$.

In order to gain insight into the properties of the system following such a quench, we will compute \emph{static} (equal time) correlation functions of the form $C_{\hat{O}}(x,t)\equiv\langle e^{iH_ft}\hat{O}(x)\hat{O}(0)e^{-iH_ft}\rangle$, where $\hat{O}$ is an operator and $\langle\ldots\rangle$ stands for the average taken over the initial state. Notice that, since in general the initial state is not an eigenstate of $H_f$, time translation invariance is broken and $C_{\hat{O}}(x,t)$ is explicitly time dependent. To begin with, let us consider the one-particle density-matrix
\begin{equation}\label{opdm}
    C_{\psi_r}(x,t)=\langle e^{iH_ft}\psi^{\dagger}_r(x)\psi_r(0)e^{-iH_ft}\rangle.
\end{equation}
whose Fourier transform leads to the instantaneous momentum distribution function. This correlation function can be calculated using the bosonization identity
\begin{equation}\label{eq:bos_id}
\psi_r(x)=F_r\frac{1}{\sqrt{2\pi a}}e^{-i\phi_r(x)},
\end{equation}
where
\begin{equation}
\phi_r(x)=-r\sum_{q\neq 0}\frac{e^{-aq/2}}{\sqrt{n_q}}(e^{irqx}b_{qr}+e^{-irqx}b^{\dagger}_{qr}),
\end{equation}
$a$ is an UV cutoff used to regularize the short distance divergences of the model and the $F_r$, usually called Klein factors, are unitary operators obeying $\{F^{\dagger}_r,F_{r'}\}=2\delta_{rr'}$. After the quench we obtain the following result for this correlation function in the thermodynamic limit (see details in Appendix \ref{app_corr}):
\begin{multline}\label{eq:corr}
C_{\psi_r}(x,t)=C^{(i)}_{\psi_r}(x)\\
\times\exp\left\{\frac{(K_i+K^{-1}_i)}{2}\Phi_1(x,t)+\frac{(K_i-K^{-1}_i)}{2}\Phi_2(x,t)\right\},
\end{multline}
where the functions
\begin{multline}\label{phi1}
\Phi_{1}(x,t)=-2 \int^{\infty}_{0} \frac{dq}{q}e^{-aq}\sinh^2(2\theta_f(q))\\
\times\sin^2(\epsilon_f(q)t)(1-\cos(qx)),
\end{multline}
and
\begin{multline}\label{phi2}
\Phi_{2}(x,t)=2 \int^{\infty}_{0} \frac{dq}{q}e^{-aq}\cosh(2\theta_f(q))\\
\times\sinh(2\theta_f(q))\sin^2(\epsilon_f(q)t)(1-\cos(qx))
\end{multline}
are independent of the initial interactions. In these equations
\begin{equation}\label{eq:corr_initial}
C^{(i)}_{\psi_r}(x)=\frac{i}{2\pi}|x|^{-1/2(K_i+K^{-1}_i)}
\end{equation}
is the correlation function at $t=0$, \textit{i.e.}, the \emph{equilibrium} equal-times correlation function for a system described by $H_i$.

\subsubsection{Steady state and the generalized Gibbs ensemble}

In order to investigate the steady state properties after a quench to a Coulomb potential, we note that for $t\rightarrow\infty$ the oscillatory, time-dependent factor within the integrals in Eqs. (\ref{phi1}) and (\ref{phi2}) average to $1/2$, regardless of the divergence in the mode velocity. By evaluating the resulting integral we obtain the long distance $x \gg d$ behavior of the one-particle density matrix, which takes the following asymptotic form:
\begin{align}\label{opdm_stat}
  C_{\psi_r}(x,\infty)\simeq C^{(i)}_{\psi_r}(x)e^{-\frac{g}{4}K_i\,\log^{2}(x/d)}.
\end{align}
We thus see that the leading order is given by the correction introduced by the quench, that decays faster than any power law, and in particular decays faster than the equilibrium correlations in the system with long-range interactions, where it has a leading behavior given by $e^{-c\log^{3/2}(x/d)}$~\cite{schulz93_wigner_crystal_1d}. In addition to the presence of $C^{(i)}_{\psi_r}(x)$ as an overall prefactor, the memory of the initial condition is also reflected by the factor $K_i$ that enters in the exponent of the correction introduced by the quench and modulates the coupling $g$.

Few-body correlations after a quantum quench in many exactly solvable models, in particular in those that can be factorized in separate momentum sectors, have been shown to relax to a generalised Gibbs ensemble (GGE)~\cite{rigol07_generalized_gibbs_hcbosons,calabrese07_quench_CFT_long,barthel08_quench,calabrese11_quantum_ising_quench}. The LM belongs to this group, even though some other quantities such as energy fluctuations cannot be obtained from the GGE~\cite{cazalilla06_quench_LL,iucci09_quench_LL}. In Ref.~\onlinecite{cazalilla12_thermalization_correlations} two main ingredients were identified as necessary to show the equivalence of the steady state static correlations with the GGE, namely, dephasing among different Fourier components and nonergodicity of the correlations, \emph{i.e.} the fact that asymptotically correlations depend only on the eigenmode occupations. Other integrable models have been shown to relax to the GGE~\cite{fioretto10_quench_integrable_theories,panfil12_lieb_liniger_out_equilibrium_luttinger}, though performing the trace associated to the GGE for specific interacting systems represents an almost impossible task. Moreover, a general result, which is always desirable, relating relaxation in integrable models to the GGE is still lacking.

In a quantum quench into the Coulomb Luttinger model, the singularity present in $v(q)$ when $q\to0$ does not affect the dephasing (since the energy remains finite for any $q$) while preserves the nonergodicity of the correlations implying therefore the equivalence with the GGE. However, the equivalence of dynamic correlations is less clear. Let us show that the equivalence holds, by considering as an example the stationary limit of the two times Green's function:
\begin{equation}\label{opdm2t}
    C_{\psi_r}(x,t,t_0)=\langle \psi^{\dagger}_r(x,t+t_0)\psi_r(0,t_0)\rangle,
\end{equation}
where, as before, $\langle\ldots\rangle$ stands for the expectation value taken over the initial state, that we will take as a non-interacting Fermi gas ($K_i=1$) for simplicity. After using the bosonization identity Eq. (\ref{eq:bos_id}) and the Baker-Hausdorff formula, we obtain:
\begin{multline}\label{opdm_dyn}
    C_{\psi_r}(x,t,t_0)=\frac{\left(1-e^{-2\pi a/L}\right)^{-1}}{L}e^{[\phi_r(x,t+t_0),\phi_r(0,t_0)]}\\
    \times\langle e^{i(\phi_r(x,t+t_0)-\phi_r(0,t_0))}\rangle,
\end{multline}
where the commutator is a c-number that only depends on $t$. Since we are dealing with a Gaussian theory, we then make use of the property $\langle e^A\rangle=e^{-\frac{1}{2}\langle A^2\rangle}$. The quantity of interest is essentially the exponent,
\begin{multline}\label{eq:dyn_green_result}
    \langle(\phi_r(x,t+t_0)-\phi_r(0,t_0))^2\rangle= \\
    \sum_{q>0}\frac{e^{-aq}}{n_q}\{|e^{iqx}f(q,t+t_0)-f(q,t_0)|^2\\
    +|e^{-iqx}g(q,t+t_0)-g(q,t_0)|^2\}
\end{multline}
which, by using the expressions for $f(q,t)$ and $g(q,t)$ given in Appendix \ref{app_formal_sol} can be shown to reduce in the limit $t_0\to\infty$ to\begin{multline}\label{eq:dyn_green_gge}
\lim_{t_0\to\infty}\langle (\phi_r(x,t+t_0)-\phi_r(0,t_0))^2\rangle=\\
2\sum_{q>0}\frac{e^{-aq}}{n_q}\times\{\cosh^2(2\theta_f(q))(1-\cos(qx)\cos(\epsilon_f(q)t))\\
-\cosh(2\theta_f(q))\sin(qx)\sin(\epsilon_f(q)t)\}.
\end{multline}
 
Next, we introduce the GGE with which we shall compare the above result. In terms of the bosonic basis that diagonalizes the evolution Hamiltonian $H_f$, $\{\alpha_q,\alpha^{\dagger}_q\}$, (see Appendix~\ref{app_formal_sol}), the first natural choice for the GGE density operator is
\begin{equation}
\rho_{\mathrm{gG}}=\frac{1}{Z_{\mathrm{gG}}}\exp\left[\sum_{q>0}\lambda(q)\hat{n}(q)\right],
\end{equation}
where $\hat{n}(q)=\alpha^{\dagger}_q\alpha_q$, $Z_{\mathrm{gG}}=\Tr\left[\exp\sum_{q>0}\lambda(q)\hat{n}(q)\right]$, and the Lagrange multipliers $\lambda(q)$ are fixed by the initial conditions:
\begin{equation}
\lambda(q)=\langle\hat{n}(q)\rangle_{\mathrm{gG}}=\langle\hat{n}(q)\rangle=\sinh^2(\theta_f(q)),
\end{equation}
where $\langle\hat{O}\rangle_{\mathrm{gG}}=\Tr[\rho_{\mathrm{gG}}\hat{O}]$. In the following we will argue that, if we define $C_{\psi_r}^{\mathrm{gG}}(x,t)\equiv\langle\psi^{\dagger}_r(x,t)\psi_r(0,0)\rangle_{\mathrm{gG}}$, then
\begin{equation}
\lim_{t_0\rightarrow\infty}C_{\psi_r}(x,t,t_0)=C_{\psi_r}^{\mathrm{gG}}(x,t).
\end{equation}
To show this, we proceed in complete analogy with the previous calculation. We obtain in the GGE the result
\begin{multline}\label{eq:gge_result}
C_{\psi_r}^{\mathrm{gG}}(x,t)=\frac{\left(1-e^{-2\pi a/L}\right)^{-1}}{L}e^{[\phi_r(x,t),\phi_r(0,0)]}\\
    \times\exp\left[-\frac{1}{2}\langle (\phi_r(x,t)-\phi_r(0,0))^2\rangle_{\mathrm{gG}}\right].
\end{multline}
Again we are only interested in the exponent, being the remaining factors equal to Eq. (\ref{opdm_dyn}). A calculation analogous to the one leading to Eq. (\ref{eq:dyn_green_gge}) shows a complete equivalence with the exponent of Eq. (\ref{eq:gge_result}).

Interestingly, it was recently shown~\cite{essler12_dynamical_gge} that the same GGE that reproduces the static (one time) correlations after relaxation can describe the dynamic (two times) correlation functions. The proof is based on the Lieb-Robinson bounds~\cite{lieb72_bound_spin_chain} on the information propagation speed in a system with \emph{short-range} interactions. The fact that in our system the dynamic correlators in the steady state can be mimicked by a GGE suggests that the short-range interaction hypothesis can be relaxed in some cases. This is related to the fact that although there is no finite speed limit for the information propagation carried by the plasmonic QPs, correlations change their behavior when crossing a non-linear light cone, as we will discuss in the next section.

\subsubsection{Dynamics and non-linear light cone effect}

\begin{figure}
  \includegraphics[width=\figwidth]{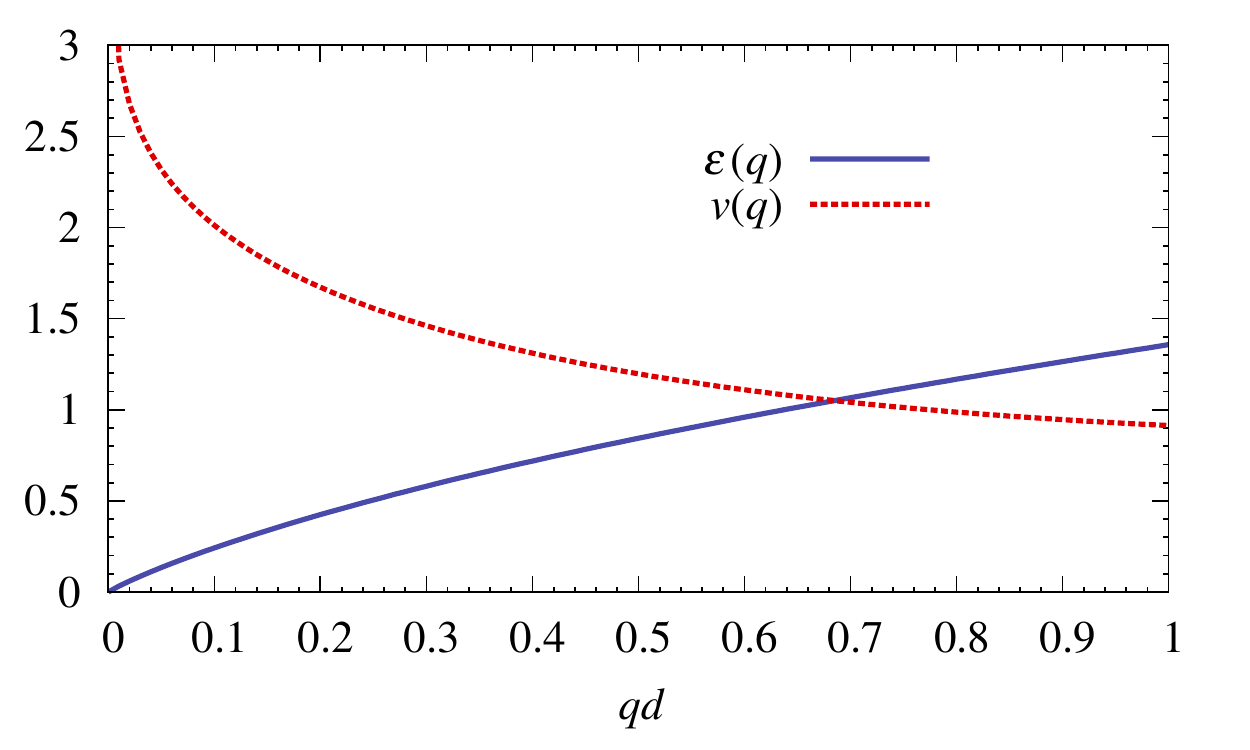}\\
  \caption{(Color Online) Spectrum $\varepsilon(q)$ (solid line) and group velocity $v(q)$(dashed line) of the bosonic QPs of $H_f$, for $g=1$. $\varepsilon(q)$ is given in units of $\frac{v_F}{d}$ and $v(q)$ in units of $v_F$.}\label{fig_spectrum}
\end{figure}

Now we turn to the subject of finite time dynamics of the system after the quench. Generally, after a quantum quench the space-time dependence of the correlations in systems with short-range interactions exhibit the so-called light-cone effect: in the region of space such that $x>2 v t$, where $v$ is the characteristic velocity of the excitations, correlations essentially keep the form already present in the initial state. In the complementary region $t>x/2 v$, they are of the form of the steady state correlations. This effect has been rigourously demonstrated for one dimensional models with conformal invariance~\cite{calabrese06_quench_CFT} of which the LM is a specific example, and can be understood as follows: in these models the dispersion relation of the QPs is linear $\epsilon(q)=v q$, with $v=\frac{d\epsilon(q)}{dq}$ the (constant) group velocity of the QPs. The initial state (a complicated excited state of the Hamiltonian $H_f$) acts as a source of QPs that propagate \emph{semiclassically} through the system with the velocity $v$. A pair of QPs emerging from the same point is entangled, and their arrival to two distant points affects the correlations initially present between them. This picture has been validated in a series of theoretical works~\cite{cazalilla06_quench_LL,iucci09_quench_LL,iucci10_quench_sine_gordon,cramer08_exact_relaxation_lattice_systems,cramer08_relaxation_superlattices,barmettler12_light_cone__bosons,lauchli08_light_cone_bose_hubbard} and even experimentally~\cite{chenau12_light_cone_experimental_bosons}. This simple interpretations faces a problem in one dimension when the interactions are long-ranged, since the group velocity $v_f(q)$ diverges for $q\rightarrow 0$ whereas the energy is finite [see Fig. (\ref{fig_spectrum})] . Moreover, the QP momentum distribution $\langle\alpha^{\dagger}_q\alpha_q\rangle=\sinh^2(\theta_f(q))$ (using the notation of the Appendix \ref{app_formal_sol}), which is a constant of motion, is peaked around $q=0$, the region where the velocity varies more abruptly. Thus, if we stick to the above described picture, we would expect that the correlations of the steady state propagate \emph{instantaneously} throughout the system. We shall see that this is not the case and point out the failure of the above reasoning when long-range interactions are present.

We first show some analytical results in the asymptotic regimes. We start by analyzing the integral defining $\Phi_1(x,t)$ in Eq. (\ref{phi1}) (being the analysis completely analogous for $\Phi_2(x,t)$) since this integral define the correction induced by the quench on top of the initial condition $C^{(i)}_{\psi_r}(x)$. After integration by parts we find that for $x\gg d$ and fixed $t$
\begin{multline}
\Phi_{1}(x,t) \simeq 2\log(dx^{-1})\sinh^{2}(2\theta_f(x^{-1}))\sin^{2}(\epsilon_f(x^{-1})t)\\
+\xi(t),\label{an_lc}
\end{multline}
where
\begin{multline}\label{xi}
\xi(t) = 2\int^{\infty}_0 \frac{dq}{q}\sinh^{2}(2\theta_f(q))\sin^2(\epsilon_f(q)t)\\
\simeq -\frac{g}{2}\log^{2}(2v_Ft/d)
\end{multline}
is a function that only depends on $t$. The last equality is valid for $v_F t\gg d$\footnote{For $v_F t\ll d$, $\xi(t)$ grows as $t^2$.}. The first term in Eq. (\ref{an_lc}) vanishes approximately as $-\frac{(v_Ft)^2}{x^2}$ for $2\epsilon_f(x^{-1})t\ll 1$, representing a subdominant correction with respect to the correlations in the initial state, Eq. (\ref{eq:corr_initial}). The same analysis holds for $\Phi_2(x,t)$. We thus see that the correction induced by the quench on the one-particle density matrix reduces, up to exponentially suppressed corrections, to a time-dependent factor:
\begin{equation}\label{eq:out_light_cone}
C_{\psi_r}(x,t)=C_{\psi_r}^{(i)}(x)e^{-\frac{g}{4}K_i\log^2(2v_Ft/d)},
\end{equation}
for times such that
\begin{equation}\label{light_cone}
\frac{d}{v_F}\ll t\ll \tilde{t}_x=\frac{x}{2v_F\sqrt{1-2g\log(dx^{-1})}}.
\end{equation}
Here we have used the explicit form of the plasmon dispersion Eq. (\ref{spectrum}), and we consider distances $x\gg d$. For such short times, spatial correlations have the same form as in the initial state, \textit{i.e.} are given by $C^{(i)}_{\psi_r}(x)$. 

Note that the denominator of Eq. (\ref{light_cone}) is essentially the group velocity $2v_f(x^{-1})$ and therefore we can rewrite Eq. (\ref{light_cone}) as $\frac{x}{2v_f(x^{-1})}\gg t$. For a short-range potential the group velocity $v_f(q)$ tends to a constant $v_f(0)$ as $q\rightarrow 0$, thus determining a linear light-cone $\frac{x}{2v(0)}=t$ for sufficiently large $x$. On the contrary, the slow group velocity divergence at $q\rightarrow 0$ of the long-range potential generates a (weak) non-linear light-``cone'' as suggested by Eq. (\ref{light_cone}), in which the behavior of the correlation function crosses over from a short times regime to a long times one. This result can be understood in terms of the following picture: the peculiarities of the 1D plasmon dispersion reflect in that for decreasing $q$, the energy of the plasmonic modes decreases whereas its velocity \emph{diverges} [see Fig. (\ref{fig_spectrum})]. Consequently, and since a mode with momentum $q$ can only propagate correlations (information) over distances \emph{larger} that $1/q$ (its wavelength), the minimum distance over which the new correlations can propagate is greater for the faster modes. Alternatively, at short times only high energy modes are active, but these have low velocity and therefore do no propagate far. At larger times, the lower energy modes that get progressively activated are faster and therefore reach more distant points. This argument is in agreement with Eq. (\ref{light_cone}). Hence, long-range interactions reveal more explicitly the quantum character of the QPs propagation. The sublinear behaviour of the light-cone also implies that quantum information can propagate faster than any finite speed of sound. However, this effect is not due to an acceleration of excitations with defined momentum, since each mode has constant velocity $v(q)$, but to the quantum nature of the collective behavior, which only activate fast modes at later times. Whether this effect will be also present in actual lattice models with long-range interactions remains an open problem. The fact that it lyes on the peculiarities of the plasmon dispersion in 1D suggest its generality. In such a case, and since deviations from linearity in the light cone are not strong but just a logarithmic correction, simulations with current algorithms should be feasible.

Cases of non-linear information propagation have also been reported in systems with short-range interactions, however, the mechanisms behind are totally different. For instance there has been detected certain type of bosonic models for which the light cone can "bent outwards" (accelerated information propagation)~\cite{eisert09_light_cone_accelerating}. In addition, it has been predicted that in disordered spin chains the light cone can "bent inwards", with the radius of the light cone growing logarithmically with time~\cite{burrell07_light_cone_disordered_spin_chains,burrell09_light_cone_fluctuating_disorder}.

\begin{figure}
  \includegraphics[width=\figwidth]{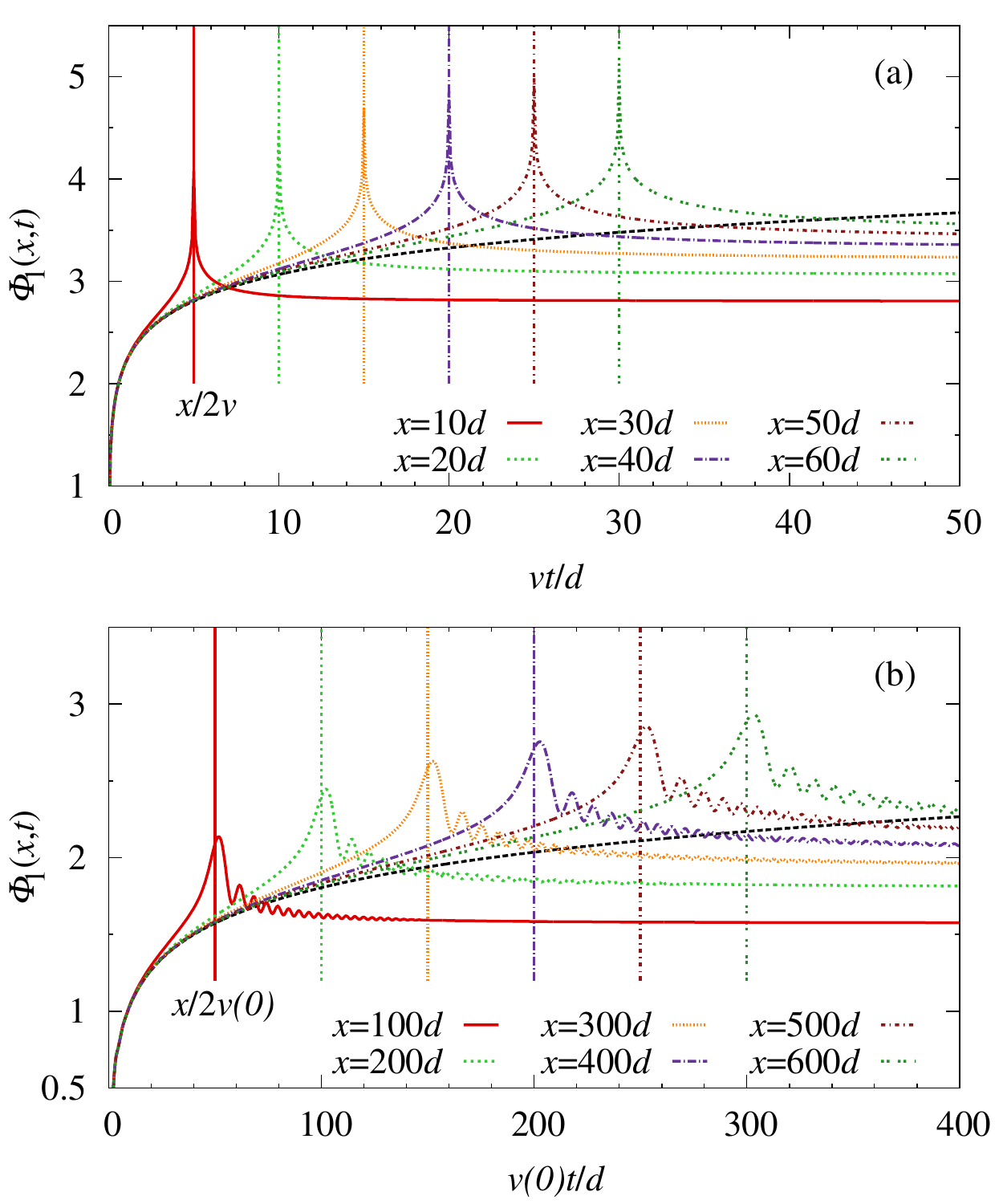}\\
  \caption{(Color Online) Time evolution of $\Phi_1(x,t)$ from numerical integration for various values of $x$ after suddenly turning on (a) a contact potential ($K_f=0.5$) and (b) a gaussian potential ($2\sigma^2=d^{-2}$). The black dashed curve is, in both graphics, the result for $x\rightarrow\infty$. The vertical lines are, for each $x$, the times defined by Eqs. (\ref{eq:light_cone_contact}) and (\ref{eq:light_cone_gauss}) respectively.}\label{fig_phi_potential}
\end{figure}

In order to complement the previous discussion we shall present some numerical calculations. First we will carefully analyze the behavior of $\Phi_1(x,t)$ in the case where we suddenly turn on short-range interactions. In Fig. (\ref{fig_phi_potential}) we show the temporal dependence of $\Phi_1(x,t)$ for different values of $x$ after turning on a contact potential
\begin{equation}
V(q)=V,
\end{equation}
and a gaussian potential
\begin{equation}
V(q)=V(0)\exp\left[-\frac{q^{2}}{2\sigma^{2}}\right].
\end{equation}
At short times, initial state correlations dominate and correlations follow the $x\rightarrow \infty$ curve. Then deviate from the asymptotic behaviour, develope a maximum and decay to the long-times value signaling the stationary state behavior. The differences between these two cases are also interesting. For the contact potential the maximum is a sharp peak (that translates into a singularity in the derivative of the correlation function) located exactly at
\begin{equation}\label{eq:light_cone_contact}
t=\frac{x}{2v},
\end{equation}
with $v=v_F\sqrt{1+2\frac{V}{\pi v_F}}$ the group velocity of the collective modes. For the gaussian potential the peak broadens and oscillations appear mounted on the final decay to the constant. The position of the peaks is approximately located at
\begin{equation}\label{eq:light_cone_gauss}
t=\frac{x}{2v(0)},
\end{equation}
where $v(q)$ is the group velocity induced by the gaussian potential, but deviations from this value are visible. This illustrates our previous argument, according to which a finite range potential is related to a linear light cone in virtue of the finiteness of the potential at $q=0$. Moreover, we can conclude that a finite dispersion in the velocity distribution of the QPs that carry the new correlations induces oscillations of the correlation functions and broadens the crossover region in which correlations change their behavior.

\begin{figure}
  \includegraphics[width=\figwidth]{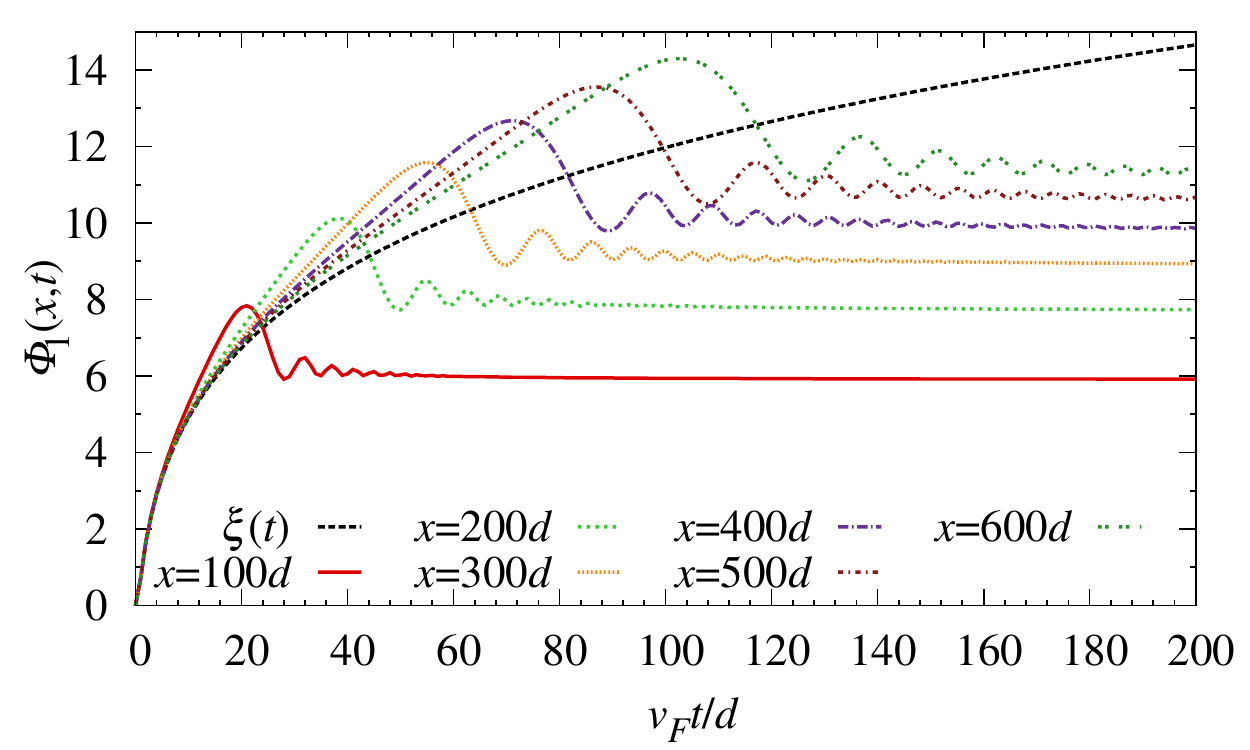}\\
  \caption{(Color Online) Time evolution of $\Phi_1(x,t)$ for various values of $x$ after a quench from a short-range interacting initial state to a long-range interacting one with $g=1$, from numerical integration. The dashed line represents the function $\xi(t)$ defined in Eq. (\ref{xi}).}\label{fig_phi1}
\end{figure}

\begin{figure}
  \includegraphics[width=\figwidth]{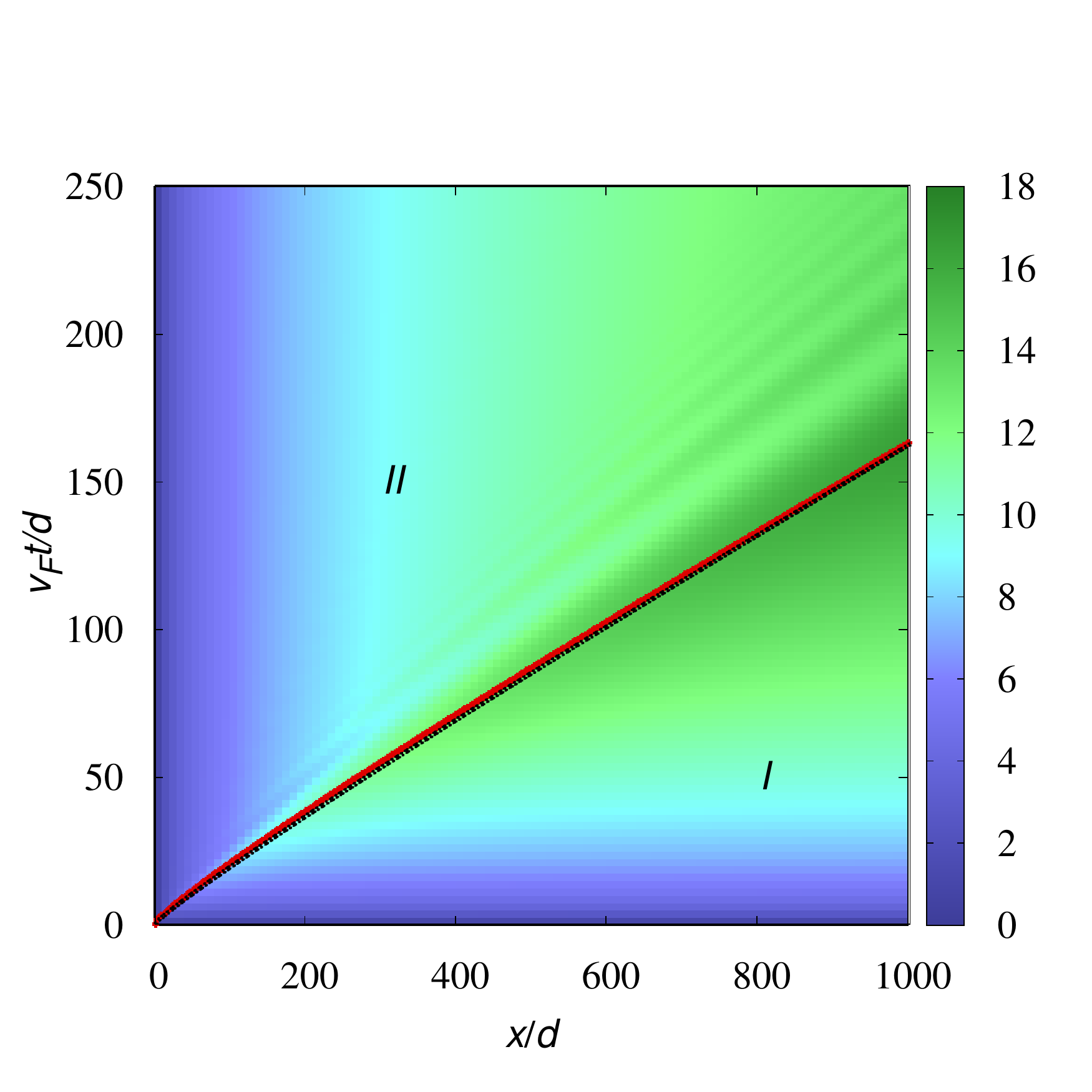}\\
  \caption{(Color Online) Density plot of the function $\Phi_1(x,t)$ for $g=1$ from numerical integration. The black dashed line is the curve $t=1.26\tilde{t}_x$ (Eq. \ref{light_cone}). The solid red line marks the positions of the maximums. Region I is the region where the initial state correlations dominate. In region II correlations have the form of the steady state.}\label{fig_light_cone}
\end{figure}

For the Coulombian potential, Fig. (\ref{fig_phi1}), some similar qualitative features are found, although in this case the peak is appreciably more broader than in the case of the gaussian potential since the velocity of the plasmons ranges over an infinite set of values. Nevertheless a much more interesting difference arises since the maximums do not lie on a straight line in the space-time plane (as was the case in the previous examples). This can be seen in Fig.~(\ref{fig_light_cone}), where we summarize the non-linear light cone effect picture. In region $I$ the correlations are the same as in the initial state up to a time dependent prefactor (Eq.~(\ref{eq:out_light_cone})) while in region $II$ the steady state correlations Eq.~(\ref{opdm_stat}) dominate. In between $I$ and $II$ there is a broad crossover region where the function reaches a maximum and oscillates. The center of the crossover region is well estimated by the position of the maximums,that is accurately described by the curve $t=1.26\tilde{t}_x$ (Eq. \ref{light_cone}). Deviations from linearity are weak since they are given by a factor $\log^{1/2} x/d$.

Finally we note that the QP dynamics has shown to be intimately related to the growth of the entanglement entropy. For example, in a CFT, the well defined propagation velocity of the QPs is directly related to the linear growth of the entanglement entropy in time after a global quench~\cite{calabrese05_evolution_entanglement_1d}. Therefore, we can speculate that the entanglement entropy growth after a global quench in the Coulomb LM will follow the same non-linear grow as found earlier for the light-cone. However, this can be only confirmed through an explicit calculation of this quantity which is out of the scope of the present article.

\subsubsection{Dynamics after a quantum quench from a non-interacting initial state}

\begin{figure}
  \includegraphics[width=\figwidth]{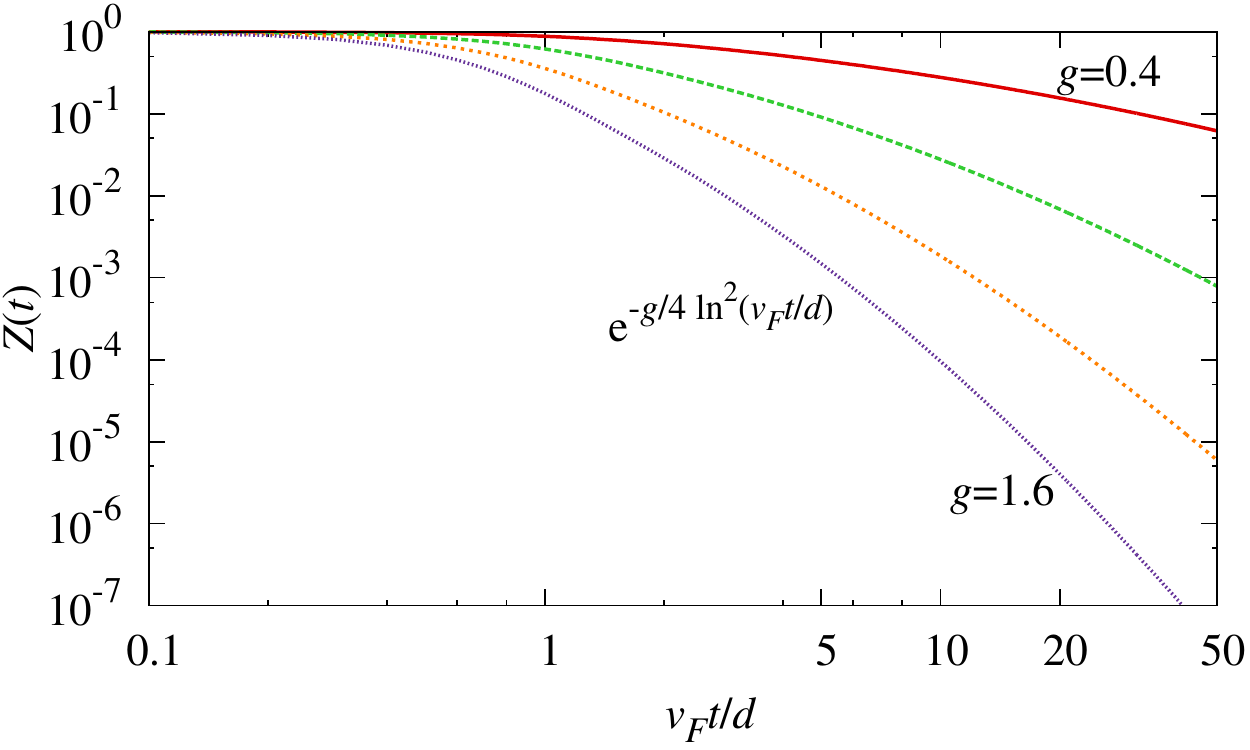}\\
  \caption{(Color Online) The Landau quasiparticle weight $Z(t)$ for different values of the coupling constant $g$, from numerical integration. Note the log-log scale; straight lines would be expected for power laws.}\label{fig_z}
\end{figure}

Let us analyze next in more detail the case of a non interacting initial state, for which Eqs.~(\ref{eq:corr})-(\ref{phi1}) hold provided that one takes $K_i=1$. We shall consider two quantities that were utilized in studying the universality of the Luttinger liquid description after an interaction quench~\cite{karrasch12_ll_universality_quench}, namely, the time evolution of the $Z$-factor jump of the instantaneous momentum distribution function $n(k, t)$ at the Fermi momentum $k_F$ and the kinetic energy per length $e_\mathrm{kin}(t)$. The discontinuity of the momentum distribution is
\begin{equation}
Z(t) = \lim_{k\rightarrow k_{F}^{+}}n(k,t)-\lim_{k\rightarrow k_{F}^{-}}n(k,t),
\end{equation}
where $n(k,t)$ is the instantaneous Fourier transform of the one-particle density matrix Eq~(\ref{opdm}). $Z(t)$ can be interpreted as a time-dependent `Landau quasiparticle" weight in an effective Fermi liquid description of the system at finite times~\cite{cazalilla06_quench_LL,iucci09_quench_LL}. For quenches involving a Hamiltonian with only finite range interactions, and thus characterized by parameters $K_f$ and $v_f$, $Z(t)\sim t^{-\frac{1}{4}(K_f^2-K_f^{-2}-2)}$~\cite{iucci09_quench_LL,rentrop12_lm_quench_momentum_dependence}. The same computation in the case of a long-range Coulomb potential yields the result
\begin{equation}
Z(t)\sim e^{-\frac{g}{4}\log^{2}(2v_Ft/d)}
\end{equation}
which decays faster than any power law but still slower than an exponential law. Thus, at least at the level of the one-particle density matrix and the momentum distribution, correlations tends to the steady state value faster for a quench involving long-range interactions than for short-range ones. The behavior of $Z(t)$ is depicted in Fig. (\ref{fig_z}).

\begin{figure}
  \includegraphics[width=\figwidth]{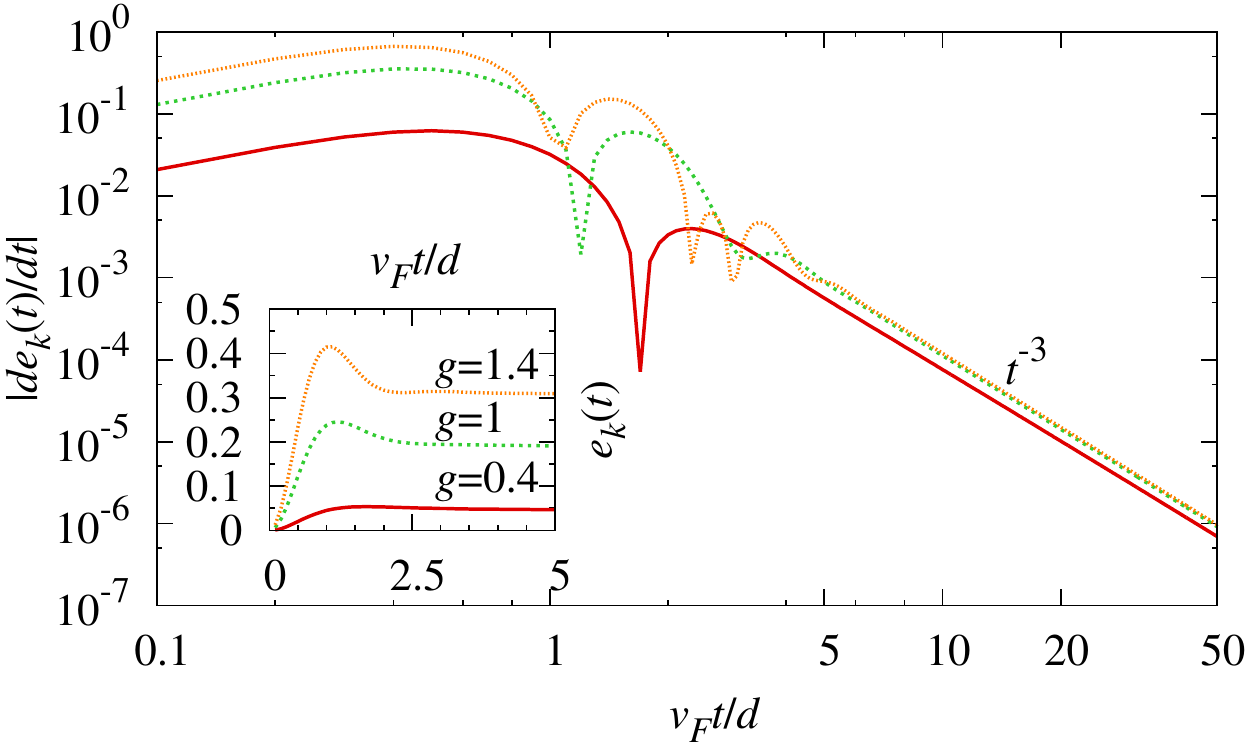}\\
  \caption{(Color Online) The derivative of the kinetic energy per unit length for different values of the coupling constant $g$, from numerical integration. In the inset we can see that the kinetic energy per unit length reaches a maximum and goes to a stationary state constant value. $e_\mathrm{kin}(t)$ is given in units of $\frac{v_F}{2\pi d^2}$.}\label{fig_dek}
\end{figure}

We shall also focus on another quantity, the kinetic energy per unit length, defined as
\begin{align}\label{ekin}
\nonumber e_\mathrm{kin}(t)&=\langle e^{iH_ft} H_i e^{-iH_ft}\rangle \\
&=\frac{v_F}{2\pi}\int_0^{\infty}dq\,\ q \sinh^2(2\theta_f(q))\sin^2(\epsilon_f(q)t).
\end{align}
It describes the rate at which the excitations of the initial state fade away. In the case of turning on finite range interactions $|\frac{de_\mathrm{kin}(t)}{dt}| \sim \frac{\frac{1}{4}(K_f^2-K_f^{-2}-2)v_F}{4\pi v_f^2}t^{-3}$~\cite{karrasch12_ll_universality_quench} in the long-times limit. For of long-range interactions, the contribution coming from the divergence of the potential only represents a logarithmic correction to the power law, and the same $t^{-3}$ law holds, as shown in Fig. (\ref{fig_dek}). We thus see that the long-range character of the interactions affects some of the aspects of the dynamics, while contributes as a subdominant correction to others. Strong evidence has been provided~\cite{karrasch12_ll_universality_quench} pointing to the universality in the Luttinger liquid sense of the asymptotic time dependence of these two quantities for short-range interactions. The extent to which this universality is maintained for long-range interactions constitutes an open problem.

\subsection{Vertex operators correlations and particle-number fluctuations}

Other correlation functions exhibit a similar behavior to the one-particle density matrix. Let us now focus on the correlators:
\begin{align}
 C_{\phi}^{m}(x,t)&=\langle e^{2im[\phi(x,t)-\phi(0,t)]}\rangle\\
 &=e^{-2m^2\langle[\phi(x,t)-\phi(0,t)]^2\rangle},\\
 C_{\theta}^{n}(x,t)&=\langle e^{in[\theta(x,t)-\theta(0,t)]}\rangle\\
 &=e^{-\frac{n^2}{2}\langle[\theta(x,t)-\theta(0,t)]^2\rangle},
\end{align}
where
\begin{equation}
\phi(x,t)=\frac{\phi_R(x,t)-\phi_L(x,t)}{2}
\end{equation}
and
\begin{equation}
\theta(x,t)=-\frac{\phi_R(x,t)+\phi_L(x,t)}{2}
\end{equation}
is the canonically conjugated field of $\phi(x,t)$, \textit{i.e.}
\begin{equation}
[\phi(x,t),\frac{1}{\pi}\partial_{x'}\theta(x',t)]=i\delta(x-x').
\end{equation}
These correlation functions are related to the total density and current-density fluctuations in the LM and in Luttinger liquids in general. In fact, in the LM we can write the fermionic particle density in real space as~\cite{haldane81_luttinger_liquid}:
\begin{equation}\label{phi_dens}
  \partial_x \phi(x,t) = -\pi [\rho_R(x,t)+\rho_L(x,t)],
\end{equation}
in the $L\to\infty$ limit.

More generally, in the context of the hydrodynamic approach of 1D interacting systems (the Luttinger liquid theory~\cite{haldane81_effective_harmonic_fluid_approach}), $C_{\phi}^{m}(x)$ and $C_{\theta}^{n}(x)$ describe the wave number and phase fluctuations near $2mk_F$ and $nk_F$, respectively, of the density correlation function $\langle\rho(x)\rho(0)\rangle$.

In the Coulomb Luttinger liquid in equilibrium the slow decay of the $4k_F$ component of the density $C_{\phi}^{2}(x)~e^{-C\log^{1/2}x}$, slower than the $2k_F$ component, hallmarks the appearance of the Wigner crystal phase. After a quantum quench from an initial state with short-range interactions, we obtain in the steady state the following correlations:
\begin{eqnarray}\label{phi_theta_on}
   C_{\phi}^{m}(x,\infty) &\simeq& C^{(i),m}_{\phi}(x)\left(x^{K_i/2}\right)^{2m^{2}},\\
    C_{\theta}^{n}(x,\infty) &\simeq& C^{(i),n}_{\theta}(x)\left(e^{-\frac{gK_i}{2}\log^{2}(x/d)}\right)^{n^{2}/2},
\end{eqnarray}
where $C^{(i),m}_{\phi}(x)=|x|^{-2m^{2}K_i}$ and $C^{(i),n}_{\theta}(x)=|x|^{-\frac{n^{2}}{2}K_i^{-1}}$ represent the initial condition. The behaviour of the $C_{\theta}^{n}$ correlator is similar to the one encountered for the Green's function. Hence, at long distances the correlations are dominated by the correction factor introduced by the quench, that decays faster than any power law, and in particular faster than in equilibrium. The initial condition is only reflected in the presence of the factor $K_i$ multiplying the coupling constant $g$. On the other hand, the correlation function $C_{\phi}^{m}$ presents more interesting features: the effect of the quench is, to leading order, only reflected in a correction of the initial condition's power-law exponent, leading to a decay of the form $C_{\phi}^{m}(x,\infty)\sim |x|^{-m^2 K_i}$ for sufficiently large distances. We notice that the consequences on $C_\phi^m$ of suddenly changing the range of the interaction are similar to the ones originated in an abrupt change on the value of the stiffness parameter $K$.

The quantity $\langle[\phi(x,t)-\phi(0,t)]^2\rangle$ is also directly related to the particle number fluctuations $\mathcal{F}(x,t)$ in a segment of size $x$. In fact, by integrating Eq. (\ref{phi_dens}), we obtain that
\begin{equation}
\mathcal{F}(x,t)\equiv\langle[\hat{N}_x-\langle\hat{N}_x\rangle]^2\rangle=\pi^2\langle[\phi(x,t)-\phi(0,t)]^2\rangle,
\end{equation}
where $\hat{N}_x$ is the operator that represents the number of particles in such region, and $\mathcal{F}(x,t)$ is the quadratic mean deviation of that particle number. Notice that $C_{\phi}^{m}(x,t)$ is the generating function of the cumulants of the (gaussian) particle number probability distribution.

The fluctuations in the number of particles of a subsystem are intimately related to its entanglement entropy and has been proposed as a natural experimental way to measure many-body entanglement~\cite{klich09_entanglement_current_fluctuations_qpc,song11_entanglement_charge_statistics_free_systems,song12_entanglement_charge_fluctuations}. However this equivalence, which turns out to be a proportionality relation, has been only rigorously proven for free fermions in equilibrium, and verified in a Luttinger liquid with short-range interactions~\cite{song12_entanglement_charge_fluctuations}. In particular for these two cases $\mathcal{F}(x)\sim \log x$. Nevertheless, the equivalence breaks down for the dynamics of these two quantities after quantum quenches. For example, after an interaction quench in a short-range Luttinger liquid (a CFT), the entanglement entropy grows linearly with time until it saturates to a value that depends linearly on the size of the subsystem~\cite{calabrese05_evolution_entanglement_1d}, while the fluctuations $\mathcal{F}(x)$ grow logarithmically with time and saturate to $\log x$. Notice that fluctuations grow in time after the quench in the same fashion as they grow, in equilibrium, with the subsystem size, \textit{i.e.} logaritmically. This is related to the conformal invariance of the theory and the fact that it posses a dynamical exponent $z=1$~\cite{levine12_full_counting_statistics_disorder}.

In the Coulomb LM in equilibrium $\mathcal{F}(x)\sim\log^{1/2}(x)$, grows slower than logaritmically suggesting that the entanglement entropy scales in a different way for systems with long-range interactions. Instead, after a sudden change in the range of the interaction, we find that the particle-number fluctuations behave as:
\begin{equation}\label{eq:fluctuations}
\mathcal{F}(x,t)\simeq \left\{
                         \begin{array}{ll}
                           K_i\log(\frac{v_Ft}{d}), & \hbox{for $t\ll\frac{x}{2v_f(x^{-1})}$;} \\
                           K_i\log(\frac{x}{d}), & \hbox{for $t\gg\frac{x}{2v_f(x^{-1})}$,}
                         \end{array}
                       \right.
\end{equation}
where $v_Ft\gg d$. They grow logarithmically with time and saturate to a value that depends logarithmically on the subsystem size, \emph{i.e.} they behave as in a quench in the short-range LM. The only manifestation of the Coulomb potential is reflected in the timescale at which $\mathcal{F}$ saturates.

\subsection{Turning off the long-range interactions}\label{toff}

Next we briefly consider the opposite situation to the one analyzed above, namely the case where initially the system is in the ground state of the LM Hamiltonian Eq. (\ref{ham_lm}) with Coulomb interactions, but the time evolution is conducted by a Hamiltonian with contact interactions, characterized by parameters $K_f$ and $v_f$. Using the general form Eq. (\ref{general_form}) adapted to this case, we find the following expression for the one-particle density matrix in the steady state at leading order:
\begin{equation}\label{opdm_off}
    C_{\psi_r}(x\gg d,t=\infty) \sim e^{-\frac{\sqrt{g}}{3\sqrt{2}}(K_f^{2}+1)\log^{3/2}(x/d)}.
\end{equation}
This result is completely equivalent to one in which the Coulomb interaction strength $g$ is suddenly changed (suitably modifying the prefactors in the exponent), since the correction factor in the long-times regime has the same form as the initial correlations. Also in the steady state, we find
\begin{eqnarray}
  C_{\phi}^{m}(x\gg d,\infty) &\sim& e^{-2m^{2}K_f^{2}\frac{\sqrt{2}}{3}g^{1/2}\log^{3/2}(x/d)}, \\
  C_{\theta}^{n}(x\gg d,\infty) &\sim& e^{-\frac{n^{2}}{6}\sqrt{2}{g}^{1/2}\log^{3/2}(x/d)}.
\end{eqnarray}
We observe that the three correlation functions have a similar asymptotic behavior for large distances and that they differ from the correlations in the ground state. In particular, we see that, remarkably, $C_{\theta}^{n}$ does not contain the parameters of $H_f$; it only depends on the initial state.

Regarding the dynamics at short times, since the Hamiltonian driving the time evolution has a linear QP spectrum, the linear light-cone effect holds exactly. In fact, for times $d\ll 2v_ft\ll x$ correlations look like those of the initial state up to a time dependent prefactor:
 \begin{eqnarray}
   C_{\psi_r}(x,t)& \simeq & C^{(i)}_{\psi_r}(x)\, e^{-\frac{(K_f^{2}-1)}{6}\sqrt{2g}\log^{3/2}(2v_ft/d)}, \\
   C_{\phi}^{m}(x,t)& \simeq & C_{\phi}^{(i),m}(x)\, e^{-2m^{2}K_f^{2}\frac{\sqrt{2}}{3}g^{1/2}\log^{3/2}(2v_ft/d)}, \\
   C_{\theta}^{m}(x,t)& \simeq & C_{\theta}^{(i),n}(x)\, e^{\frac{n^{2}}{\sqrt{2}3}g^{1/2}\log^{3/2}(2v_ft/d)}.
 \end{eqnarray}
The initial condition is given by the equal-times correlation functions~\cite{schulz93_wigner_crystal_1d},
\begin{align}
C^{(i)}_{\psi_r}(x)&\sim x^{-1} e^{-\sqrt{\frac{2g}{3}}\log^{3/2}(x/d)},\\
C_{\phi}^{(i),m}(x)&\sim e^{-2m^2\sqrt{\frac{2}{g}}\log^{1/2}(x/d)},\\
C_{\theta}^{(i),n}(x)&\sim e^{-\frac{n^2}{3}\sqrt{2g}\log^{3/2}(x/d)}.
\end{align}
We notice that, as pointed out in the discussion on general quenches in Appendix \ref{app_corr}, if we completely turn off the interactions ($K_f=1$) the one-particle density matrix has no time-dependence.

Finally, the charge fluctuations follow the behavior:

\begin{equation}
\mathcal{F}(x,t)\simeq \left\{
                         \begin{array}{ll}
                           g^{1/2}K^2_f\log^{3/2}(\frac{v_ft}{d}), & \hbox{for $t\ll\frac{x}{2v_f}$;} \\
                           g^{1/2}K^2_f\log^{3/2}(\frac{x}{d}), & \hbox{for $t\gg\frac{x}{2v_f}$,}
                         \end{array}
                       \right.
\end{equation}
where $v_ft\gg d$. Notice that the fluctuations grow in time in a complete different fashion (faster) as they grow in a quench starting from a short-range interacting initial state (logarithmically). The saturation value is also different. This reflects the strong influence of the initial condition on the subsequent dynamics of the charge fluctuations.

\section{Experimental realization}\label{exp}

The experimental study of the out-of-equilibrium dynamics of isolated quantum systems has become a reality thanks to the advent of systems of ultracold atomic gases loaded on optical lattices. This systems offer the possibility of monitoring the dynamics of a quantum system that maintain coherence for long times compared to the typical experiment durations as well as a high degree of tunability of the geometric configuration and the statistics of the particles (fermions, bosons or mixtures). Nevertheless in the case of neutral atomic species, the interactions between the particles, although having a strength that is easily modified through Feschbasch resonances, are only of short-range (at most, they are of dipolar type, and decay as $1/x^3$). In this section we propose ultracold ions trapped in electromagnetic cavities as a system suitable for the study of non-equilibrium dynamics of one dimensional systems of particles coupled by the Coulomb interaction. These systems have been intensely studied in the context of experimental quantum computation, since they satisfy the basic criteria for the implementation of a quantum computer~(basically, they have long coherence times, the universal operations needed to implement any quantum algorithm can be realized using lasers tightly focused on the trapped ions, and their internal state can be accurately measured)~\cite{cirac95_ions_qc_idea,blatt12_simulations_ions,blatt08_ions_entangled,haffner08_ions_report,jane03_ions_simulation_dynamics}.

Ions can be cooled down to their ground state using laser cooling~\cite{diedrich89_laser_cooling_ions} and loaded on linear (1D) traps, commonly called linear Paul traps.  In the case of quantum computing experiments, the most popular setups use the ions $_{4}^{9}\mathrm{Be}^{+}$ (boson) or $_{20}^{40}\mathrm{Ca}^{+}$ (fermion)~\cite{haffner08_ions_report}, 
but a large variety of ions can be used. In these devices, a radio-frequency potential is applied to two electrodes which are parallel to the axis of the trap. These electrodes create an oscillating two-dimensional quadrupole potential that is translational invariant along the trap axis. If the frequency of the radio-frequency field is sufficiently large, the ions experience an effective restoring force to the center axis. Additionally, static electric fields confine the ions along the trap axis. If the confinement perpendicular to the trap axis (radial direction) is much larger than the confinement along the trap axis, cold ions form a linear crystal~\cite{raizen92_linear_cluster_coulomb_ions}, in which the spacings are determined by a balance between the horizontal (axial) confining fields and mutual Coulomb repulsion.

In the quantum computation setups\footnote{See Ref.~\cite{haffner08_ions_report} and references therein.} the measurements aim to the determination of the internal state of the qubit encoded by each ion. However, nothing prevents us from using the battery of techniques developed in the field of ultracold neutral atoms, such as the time-of-flight measurements, through which observables such as the momentum distribution and other few-point correlations are accessible on real time.

\begin{figure}
  \includegraphics[width=\figwidth]{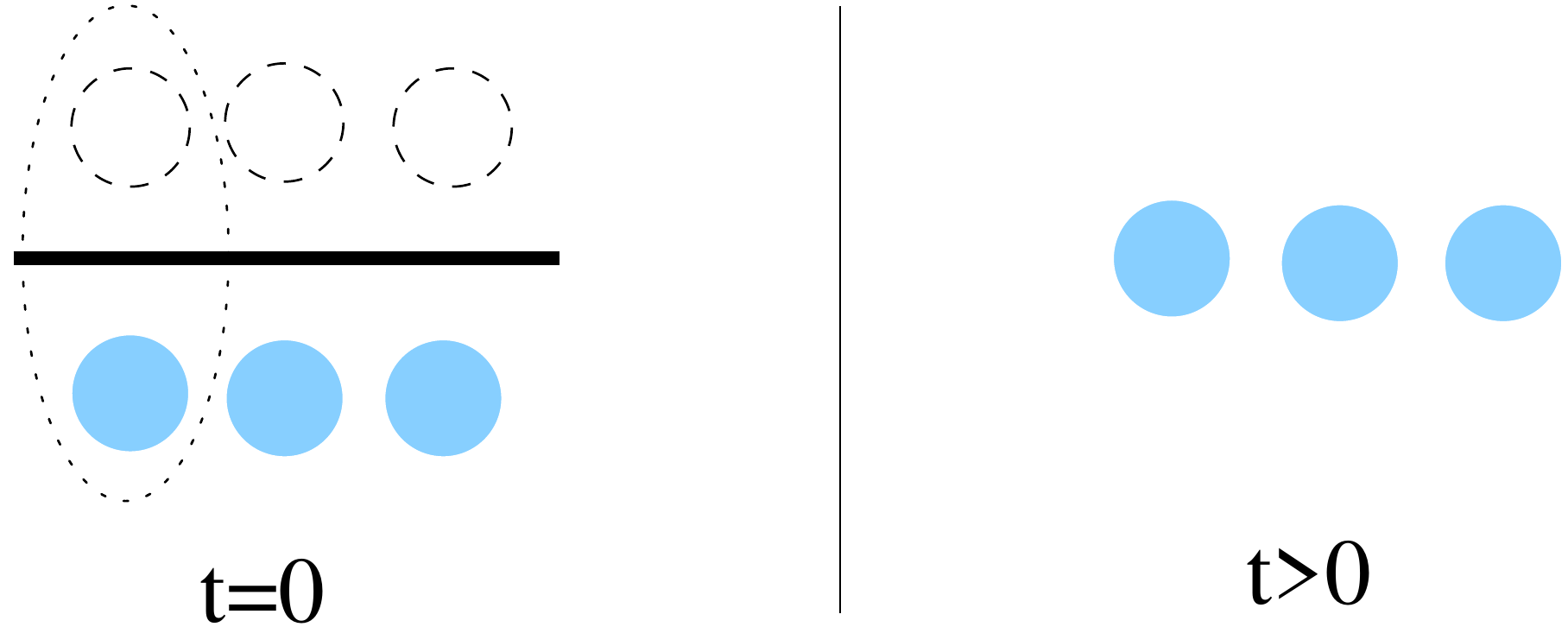}\\
  \caption{(Color Online) Scheme of the experimental realization. Initially the ions (blue) form dipoles with its image charges (dashed circles) placed symmetrically with respect to the gate (thick solid line). The dipoles interact with each other through the dipole interaction, decaying as $x^{-3}$. After suddenly removing the gate the ions interact via the bare Coulomb repulsion.}\label{fig_experiment}
\end{figure}

The particular quench sequence proposed in this paper can be realized loading the ions on the linear trap in the presence of a gate parallel to the trap axis (see Fig. (\ref{fig_experiment})). The interactions between the ions would be screened by the image charges generated by the gate, symmetrically placed with respect to it, leaving us with an effective dipole interaction between the atoms, in which case the system can be described as a short-range interacting Luttinger liquid. The sudden quench can be realized by rapidly removing the gate, allowing the ions to interact via the bare Coulomb repulsion. To reproduce exactly the sudden quench limit the gate should be removed in a time much shorter than any time scale of the system, or, equivalently, faster than any velocity scale. For a system of cold trapped $_{20}^{40}\mathrm{Ca}^{+}$ ions, the Fermi velocity can be estimated from the typical densities~\cite{raizen92_linear_cluster_coulomb_ions}, and has an order of magnitude of $1 \textrm{m\,s}^{-1}$, but this can be reduced using heavier ions. However, based on the analysis made in Ref. ~\onlinecite{dora11_ll_quench_adiabatic_to_sudden}, we can expect the sudden quench results to be valid for more general quench protocols, at least in some specific space-time regions.

\section{Summary}\label{sec:conclusions}

We have presented a detailed study of the behavior of the correlations following a special quench protocol in the LM, namely, to suddenly change the range of the interactions between the fermions from short to long-range and vice versa. We have found that the stationary state correlations do not obey the typical power law decays, even in the case of turning off the Coulomb potential. This behavior can be obtained from a suitable defined GGE, which turns out to describe both static and dynamic correlations. Moreover, the Coulomb potential modifies qualitatively and quantitatively some aspects of the dynamics while leaves other intact with respect to the case of a short-range interaction quench. In particular, it substantially changes the light-cone effect picture, by introducing non linearities that mimic an acceleration of the excitations, and thus opening the possibility of a faster quantum information propagation. This effect can be understood based on the special characteristics of the plasmonic QP spectrum in 1D, in particular the fact that the faster modes are the less energetic ones. Finally, we have proposed that linear chains of ultracold trapped ions would provide a suitable setup for studying the effects described in this work.

\acknowledgments
We thank Miguel A. Cazalilla for enlightening discussions. This work was partially supported by CONICET (PIP 0662), ANPCyT (PICT 2010-1907) and UNLP (PID X497), Argentina.

\appendix

\section{Formal solution}\label{app_formal_sol}

In this Appendix we provide the solution to a general quench in the LM. We start by defining $\{a_q,a^{\dagger}_{-q}\}$ ($\{\alpha_q,\alpha^{\dagger}_{-q}\}$) as the diagonal basis of $H_i$ ($H_f$). Since we shall work in the Heisenberg picture, the solution to the problem of the system's evolution can be expressed, at least formally, giving the time dependence of the operators $\{a_q,a^{\dagger}_{-q}\}$. We will use the explicit form of the Bogoliubov transformations relating these operators to the free basis $\{b_q,b^{\dagger}_{-q}\}$ defined in section \ref{sec:model}:
\begin{equation}\label{aop}
\left(
  \begin{array}{c}
    a_q \\
    a^{\dagger}_{-q} \\
  \end{array}
\right)=\mathcal{M}^{(i)}\left(
  \begin{array}{c}
    b_q \\
    b^{\dagger}_{-q} \\
  \end{array}\right),
\end{equation}
and
\begin{equation}
\left(
  \begin{array}{c}
    \alpha_q \\
    \alpha^{\dagger}_{-q} \\
  \end{array}
\right)=\mathcal{M}^{(f)}\left(
  \begin{array}{c}
    b_q \\
    b^{\dagger}_{-q} \\
  \end{array}\right),
\end{equation}
with
\begin{equation}
\mathcal{M}^{(l)}=\left(
          \begin{array}{cc}
            \cosh(\theta_l(q)) & \sinh(\theta_l(q)) \\
            \sinh(\theta_l(q)) & \cosh(\theta_l(q)) \\
          \end{array}
        \right)
\end{equation},
where $l=i,f$. The transformation relating the initial and final basis is thus:
\begin{gather}
\nonumber \left(
  \begin{array}{c}
    a_q \\
    a^{\dagger}_{-q} \\
  \end{array}
\right)=\mathcal{M}^{(i)}\left(\mathcal{M}^{(f)}\right)^{-1}\left(
  \begin{array}{c}
    \alpha_q \\
    \alpha^{\dagger}_{-q} \\
  \end{array}
\right)\\
=\left(\begin{array}{cc}
            \cosh(\theta_f(q)-\theta_i(q)) & \sinh(\theta_f(q)-\theta_i(q)) \\
            \sinh(\theta_f(q)-\theta_i(q)) & \cosh(\theta_f(q)-\theta_i(q)) \\
          \end{array}
        \right)\left(\begin{array}{c}
    \alpha_q \\
    \alpha^{\dagger}_{-q} \\
  \end{array}
\right).
\end{gather}

Since the $\{\alpha_q,\alpha_{-q}^{\dagger}\}$ operators have trivial time dependence, \emph{i.e.} $\alpha_q(t)=\alpha_q\exp[-i\epsilon_f(q)t]$, we arrive at the solution:
\begin{equation}\label{formal_sol}
    a_q(t) = f(q,t) a_q + g^*(q,t)a^{\dagger}_{-q}
\end{equation}
where
\begin{align}\label{formal_sol1}
    f(q,t) = \cos(\epsilon_f(q)q t)-i\sin(\epsilon_f(q) t)\cosh[2(\theta_f(q)-\theta_i(q))],
\end{align}
\begin{align}\label{formal_sol2}
    g(q,t) =  i\sin(\epsilon_f(q)q t)\sinh[2(\theta_f(q)-\theta_i(q))].
\end{align}

\section{General form of the correlation functions after an arbitrary interaction quench}\label{app_corr}

We provide here a general formula for all the correlation functions studied in this work in the case of an arbitrary quench. Defining as before:
\begin{equation}\label{co}
    C_{\hat{O}}(x,t)\equiv\langle e^{iH_ft}\hat{O}(x)\hat{O}(0)e^{-iH_ft}\rangle,
\end{equation}
we find that
\begin{equation}\label{general_form}
    C_{\hat{O}}(x,t)=C^{(0)}_{\hat{O}}(x)\exp[-\Phi_{\hat{O}}(x,t)],
\end{equation}
with

\begin{equation}
\begin{split}
\Phi_{\hat{O}}(x,t)&=-2\sum_{q>0} \frac{e^{-aq}}{n_q}\gamma_{\hat{O}}(q)\sinh[2(\theta_{f}(q)-\theta_{i}(q))]\\
& \times(1-\cos(qx))\sin^{2}(\epsilon_{f}(q) t).
\end{split}
\end{equation}
The function $\Phi_{\hat{O}}(x,t)$ defines the correction introduced by the quench, $C^{(0)}_{\hat{O}}(x)$ is the correlation function at $t=0$, \textit{i.e.} the \emph{equilibrium} equal-times correlation function for a system described by $H_i$, and
\begin{equation}\label{gamma_cases}
    \gamma_{\hat{O}}(q)=\begin{cases}\sinh(2\theta_{f}(q)) & \mbox{if } \hat{O}=\psi_r(x) \\ 2m^2\exp[-2\theta_{f}(q)] & \mbox{if } \hat{O}=e^{2im\phi(x)}\\-\frac{n^2}{2}\exp[2\theta_{f}(q)] & \mbox{if } \hat{O}=e^{in\theta(x)}, \end{cases}
\end{equation}
is a function depending only on the operator considered. It is worth noticing that if $\gamma_{\hat{O}}(q)=0$ the correlation function has no temporal dependence, retaining the same spatial dependence as in the initial state. For example the one-particle density matrix and accordingly the momentum distribution function are unaffected when the interactions are suddenly turned-off, since in that case $\sinh(2\theta_{f}(q))=0$~\cite{rentrop12_lm_quench_momentum_dependence}.

We shall illustrate this result following in some detail the calculation of $C^{m}_{\phi}(x,t)$. The case of $C^{n}_{\theta}(x,t)$ is completely analogous and the case of the Green's function has been already analyzed in detail in Ref.~\onlinecite{iucci09_quench_LL}. The mode decomposition of the $\phi(x)$ field is
\begin{equation}
\phi(x)=\frac{i}{2}\sum_{q>0}\frac{e^{-aq/2}}{\sqrt{n_{q}}}[e^{iqx}(b_{q}+b_{-q}^{\dagger})-e^{-iqx}(b_{-q}+b_{q}^{\dagger})],
\end{equation}
and using Eq. (\ref{formal_sol}) to introduce the time dependence, we find that
\begin{widetext}
\begin{equation}
\langle[\phi(x,t)-\phi(0,t)]^{2}\rangle=\sum_{q>0}\frac{e^{-aq}}{n_{q}}e^{-2\theta_f(q)}(1-\cos(qx))\{1+2\sinh^2(\theta_f(q)-\theta_i(q))+\sinh[2(\theta_f(q)-\theta_i(q))]\cos(2\epsilon_f(q)t)\}.
\end{equation}
If we separate the sum isolating a time independent part and a time dependent one that vanishes at $t=0$ we find the following factorization for $C^{m}_{\phi}(x,t)$:
\begin{equation}\label{phi_example}
C_{\phi}^{m}(x,t)=C_{\phi}^{(0),m}(x)\exp\{-4m^{2}\sum_{q>0}\frac{e^{-aq}}{n_{q}}e^{-2\theta_f(q)}(1-\cos(qx))\sinh[2(\theta_f(q)-\theta_i(q))]\sin^{2}(\epsilon_{q}t)\},
\end{equation}
\end{widetext}
where $C_{\phi}^{(0),m}(x)=\exp\{-2m^{2}e^{-2\theta_i(q)}\sum_{q>0}\frac{e^{-aq}}{n_{q}}(1-\cos(qx))\}$ is exactly the equilibrium correlator in the ground state of $H_i$, verifying the general result.


\end{document}